\documentclass[aps,pra,preprint,groupedaddress,showpacs]{revtex4}

\usepackage{graphicx}
\usepackage{dcolumn}
\usepackage{bm}
\usepackage{verbatim}
\usepackage{amsmath,amssymb}
\usepackage{relsize,exscale}
\usepackage{color}

\begin{document}
\title{Emergence of classical behavior from the quantum spin}
\author{M. Radonji\'c}
\author{S. Prvanovi\'c}
\author{N. Buri\'c}
\email[]{buric@ipb.ac.rs}
\affiliation{Institute of Physics, University of Belgrade,
Pregrevica 118, 11080 Belgrade, Serbia}

\begin{abstract}
Classical Hamiltonian system of a point moving on a sphere of
fixed radius is shown to emerge from the constrained evolution of
quantum spin. The constrained quantum evolution corresponds to an
appropriate coarse-graining of the quantum states into equivalence
classes, and forces the equivalence classes to evolve as single units
representing the classical states. The coarse-grained quantum spin
with the constrained evolution in the limit of the large spin
becomes indistinguishable from the classical system.
\end{abstract}


\maketitle

\section{Introduction}

It is generally agreed that all systems in the Nature are
described by appropriately formulated quantum theory. Therefore,
the fact that some systems, usually of macroscopic size, can be
described by radically different theory, namely the classical
physics, requires an explanation. Clarifying the meaning in which
the behavior described by the classical physics emerges from the
quantum substrate is one of the main topics of the theory of
quantum to classical relation (QCR).

Multifarious aspects of the problem of QCR have been analyzed since
the early formulations of quantum mechanics. The relevant
literature on the topic is vast and we shall single out as
illustrative examples only few reviews: The relevance of QCR to
the notorious problem of quantum measurement is often  discussed
in the papers collected in \cite{book-meas} and in \cite{Balian} which
contains references to the more recent developments. Some of the more
formal mathematical aspects of QCR are treated in \cite{Landsman}.
Dynamical aspects of QCR have been intensively discussed within
the framework of the correspondence principle and the semi-classical
methods for classically chaotic systems \cite{Hake}. Putative physical
mechanisms and the appropriate ontological considerations underlying
QCR are discussed from different points of view for example in
\cite{Decoh1,Decoh2,GRW,Ian,Brukner1}. During the last couple of
decades detailed experimental studies of the problems related to
QCR have been performed (see for example
\cite{Experiments,Experiments1,Experiments2}).

It has been realized many times that quantum and classical systems
are related by some sort of coarse-graining. The coarse-graining
enters differently in different theories of QCR, and is not always
equally strongly emphasized. In the theories of decoherences
\cite{Decoh1,Decoh2} the emphasis is on the influence of the
environment, but the description of the environment must be
coarse-grained to fulfill the desired decoherence effects. On the
other hand, authors like \cite{Jauh} and \cite{Brukner1,Brukner2},
to mention just a few representatives of the approach from two
different periods and background, emphasize the primary role of
the coarse-graining, associated with limited precision of the devices
used to observe the quantum system.

In this paper we shall analyze the role of an appropriate coarse-graining
for the emergence of a class of classical Hamiltonian systems
characterized by the spherical phase space. The points of the phase space
are parameterized by the spherical angles $(\theta,\phi)$ or by the Cartesian
coordinates $(J_x,J_y,J_z)$ constrained by $J_z^2+J_y^2+J_z^2=J^2$ and the
classical Hamiltonian is a smooth function of
$J_x(\theta,\phi),J_y(\theta,\phi),J_z(\theta,\phi)$. Our goal is to
show how the classical systems on the sphere can be derived from
the quantum spin of size $J$, i.e.\ quantum systems with $su(2)$
dynamical algebra and $(2J+1)${-dimensional} Hilbert space of states.
We shall see that the derivation of the classical system is done
in two independent steps, both of which are necessary. The first
step consist of an appropriate coarse-graining which introduces
the classical phase space. Classical like Hamiltonian system is
then defined on this phase space by appropriately constraining the
quantum Schr\"odinger evolution. The second step is the macroscopic
limit applied on this coarse-grained and constrained system. As
the result the coarse-grained and constrained system becomes
indistinguishable from a classical system on the sphere.

The same ideas have been recently utilized to study the appearance
of a classical system of nonlinear oscillators from the
corresponding quantum system \cite{us}. The two examples suggest
formulation of a general procedure which shall be briefly
discussed.

The paper is organized as follows. In the next Section we
recapitulate the geometric Hamiltonian formulation of quantum
mechanics and of constrained quantum dynamics, with the special
emphasis on the system with $su(2)$ dynamical algebra, i.e.\ the
quantum spin. This representation is used in Sec.\ III to construct
the classical model with the same dynamical algebra.
Section IV discusses the appearance of the classical system from the
macro-limit of the coarse-grained and constrained quantum system.
In Sec.\ V we summarize our presentation.

\section{Hamiltonian formulation of constrained quantum dynamics}

Investigations of the relations between classical and quantum
mechanics are facilitated if both theories are expressed using
similar mathematical language. Geometric Hamiltonian formulation
[16-25] and the geometric theory of coherent states
\cite{Perelomov} are two such representations of quantum mechanics
which are formulated in terms of mathematical objects typical for
classical Hamiltonian mechanics. In this section we shall briefly
recapitulate the geometric Hamiltonian formulation for the case of
a quantum system with finite-dimensional Hilbert space since this
will be the main tool of our analyzes. In particular we shall
summarize recently introduced description of constrained quantum
systems \cite{JaAnnPhys,Brody1,Brody2} within this formalism.
Group-theoretical and geometric treatment of the generalized
coherent states, as it was introduced by Perelomov
\cite{Perelomov}, shall be used when needed without prior
recapitulation.

\subsection{Hamiltonian framework for quantum systems}

Schr\"odinger dynamical equation on a separable and complete Hilbert
space ${\cal H}$ generates a Hamiltonian dynamical system on an
appropriate symplectic manifold. The symplectic structure, which
is needed for the Hamiltonian formulation of the Schr\"odinger
dynamics, is provided by the imaginary part of the unitary scalar
product on ${\cal H}$. In fact the Hilbert space ${\cal H}$ is
viewed as a real manifold ${\cal M}$ with a complex structure,
given by a linear operator $\mathcal{J}$ such that $\mathcal{J}^2=-1$.
If ${\cal H}$ is finite $n-$dimensional then ${\cal M}\equiv\mathbb{R}^{2n}$,
but in general ${\cal M}$ is an infinite dimensional Euclid manifold. In
what follows we shall consider only the finite-dimensional Hilbert
spaces since the irreducible representations of the spin $J$, i.e.\
the $su(2)$ algebra, are of finite dimension: $n\equiv \mathrm{dim}
{\cal H}=2J+1$. Real coordinates $(x_i,y_i)$ of a
point $\psi\in {\cal H}\equiv {\cal M}$ are introduced using
expansion coefficients $c_i$ in some basis
$\{|i\rangle,\>i=1,2,\dots,n\}$ of ${\cal H}$ as follows
\begin{subequations}
\label{e:psi}
\begin{align}
|\psi\rangle=&\sum_i c_i|i\rangle,\quad c_i=\frac{x_i+i y_i}{\sqrt{2}},\\
x_i=\sqrt{2}\,\mathrm{Re}(c_i)&,\quad
y_i=\sqrt{2}\,\mathrm{Im}(c_i),\quad i=1,2,\dots,n
\end{align}
\end{subequations}

The real manifold ${\cal M}=\mathbb{R}^{2n}$ has Riemannian and
symplectic structure. Since ${\cal M}$ is real it is natural to
decompose the unitary scalar product on ${\cal H}$ into it's real
and imaginary parts
\begin{equation}
\langle\psi_1|\psi_2\rangle=\frac{1}{2\hbar}g_{\cal M}(\psi_1,\psi_2)
+\frac{i}{2\hbar}\omega_{\cal M}(\psi_1,\psi_2).
\end{equation}
It follows that $g_{\cal M}$ is Riemannian metric on ${\cal M}$
and that $\omega_{\cal M}$ is symplectic form on ${\cal M}$.  Thus
the manifold ${\cal M}$ associated with the Hilbert space ${\cal
H}$ can be viewed as a phase space of a Hamiltonian dynamical
system, additionally equipped with the Riemannian metric which
reflects its quantum origin. A vector from ${\cal H}$, associated
with a pure quantum state, is represented by the corresponding
point in the phase space ${\cal M}$. We shall denote the point
from ${\cal M}$ associated with the vector $|\psi\rangle$ by $X_{\psi}$.

In the coordinates $(x_i,y_i)$ the Riemannian and the
symplectic structures of ${\cal M}$ are given by
\begin{equation}
g_{\cal M}=\begin{pmatrix}\mathbf{1}&\mathbf{0}\cr
\mathbf{0}&\mathbf{1}\end{pmatrix},
\end{equation}
\begin{equation}
\omega_{\cal M}=\begin{pmatrix}\mathbf{0}&\mathbf{1}\cr
-\mathbf{1}&\mathbf{0}\end{pmatrix},
\end{equation}
where $\mathbf{0}$ and $\mathbf{1}$ are zero and unit matrices of
dimension equal to the dimension of the Hilbert space. Thus, coordinates
$(x_i,y_i)$ represent canonical coordinates of a Hamiltonian dynamical
system. Consequently, the Poisson bracket between two functions $F_1$
and $F_2$ on ${\cal M}$ corresponding to the symplectic form
$\omega_{\cal M}$
is in the canonical coordinate $(x_i,y_i)$ representation given by
\begin{equation}
\{F_1,F_2\}_{\cal M}=\sum_i ({\partial F_1\over\partial
x_i}{\partial F_2\over\partial y_i}-{\partial F_2\over\partial
y_i}{\partial F_1\over\partial x_i}).
\end{equation}

A one parameter family of unitary transformations on ${\cal H}$
generated by a self-adjoined operator ${\hat H}$ is represented on
${\cal M}$ by a flow generated by the Hamiltonian vector field
$\omega_{\cal M}(-\mathcal{J}\hat{H}\psi,\cdot)=(dH)(\cdot)$ with the
Hamilton's function given by
\begin{equation}
H(X_{\psi})=\langle\psi|\hat H|\psi\rangle.
\end{equation}
Thus, quantum observables $\hat H$ are represented by functions of
the form $\langle\hat H\rangle_\psi$. Such and only such Hamiltonian
flows with the Hamilton's  function of the form (6) generate also
isometries of the Riemannian metric $g_{\cal M}$. More general Hamiltonian
flows on ${\cal M}$, corresponding to the Hamilton's function which are
not of the form (6), do not generate isometries and do not have the
physical interpretation of quantum observables.

It can be seen easily that
\begin{equation}
\{H_1,H_2\}_{\cal M}=\frac{1}{i\hbar}\langle[\hat H_1,\hat H_2]\rangle.
\end{equation}
In the remaining text we will take $\hbar=1$.
The Schr\"odinger evolution equation
\begin{equation}
|\dot\psi\rangle=-i\hat H|\psi\rangle
\end{equation}
is equivalent to the Hamilton's equations on ${\cal M}$
\begin{equation}
\dot X_{\psi}^a=\omega^{ab}\nabla_b  H(X_{\psi}).
\end{equation}
In the canonical coordinates $(x_i,y_i)$ the Schr\"odinger
evolution is given by
\begin{equation}
\dot x_i={\partial H\over\partial y_i},\qquad \dot y_i=
-{\partial H\over\partial x_i}.
\end{equation}

We have constructed the Hamiltonian dynamical system
corresponding to the Schr\"odinger evolution equation on ${\cal
H}$. In fact phase invariance and arbitrary normalization of the
quantum states imply that the proper space of pure quantum states
is not the Hilbert space used to formulate the Schr\"odinger
equation, but the projective Hilbert space. This also is a K\"ahler
manifold and can be used as a phase space of a geometrical
Hamiltonian formulation of quantum mechanics. Nevertheless, we
shall continue to use the formulation in which points of the
quantum phase space are identified with the vectors from ${\cal H}$
since it is sufficient for our main purpose.

\subsection{Constrained quantum dynamics}

The Hamiltonian framework for quantum dynamics enables one to
describe the evolution of a dynamical system generated by the
Schr\"odinger equation with quite general additional constraints
\cite{JaAnnPhys,Brody1,Brody2}. Suppose that the evolution given
by the Hamiltonian $H$ is further constrained onto a submanifold
$\Gamma$ of ${\cal M}$ given by a set of $k$ independent functional
equations
\begin{equation}\label{e:Constr}
f_l(X)=0, \> l=1,2,\dots,k.
\end{equation}
Equations of motion of the constrained system are in general
obtained using the method of Lagrange multipliers. In the
Hamiltonian form, developed by Dirac \cite{Dirac,Klader,Arnold1},
the method assumes that the dynamics on $\Gamma$ is determined by
the following set of differential equations
\begin{equation}
\dot X=\omega_{\cal M}(\nabla X,\nabla
H_{tot}), \qquad H_{tot}=H+\sum_{l=1}^k \lambda_lf_l,
\end{equation}
that should be solved together with the equations of the
constraints (\ref{e:Constr}). For notational convenience we do not
indicate in the gradient $\nabla$ that it is defined on ${\cal
M}$. The Lagrange multipliers $\lambda_l$ are functions on ${\cal
M}$ that are to be determined from the following, so called
compatibility, conditions
\begin{equation}\label{e:fdot}
\begin{split}
0=\dot f_l&=\omega_{\cal M}(\nabla f_l,\nabla H_{tot})\\&=
\omega_{\cal M}(\nabla f_l,\nabla H)+\sum_{m=1}^k\lambda_m
\omega_{\cal M}(\nabla f_l,\nabla f_m)
\end{split}
\end{equation}
on the constrained manifold $\Gamma$.  We shall not go into the
details of the standard Dirac's procedure that emphasizes the
distinction between the first and the second class constraints. In
order to apply the standard procedure, the constraints have to be
regular. A set of constraints is irregular if there is at least
one such that the derivative of the constraint with respect to at
least one of the coordinates is zero in at least one point on the
constrained manifold. Otherwise the constraints are regular. In
our case of finite-dimensional ${\cal M}$ constraints are regular
if for all $l$
\begin{equation}\label{e:Reg}
{\partial f_l\over \partial x_i}\neq 0,\quad {\partial f_l\over \partial y_i}\neq 0,
\end{equation}
for all $i,j=1,2\dots n$ and everywhere on the constrained
manifold $\Gamma$. If this is not satisfied the Dirac's
classification into the first and the second class is blurred and
the straightforward application of Dirac's recipe is not possible.
It will turn out that the case of interest here involves precisely
irregular constraints that must be described in the most
convenient way. However, if the constrained manifold is symplectic
then the Dirac's procedure of constructing the constrained system
and reducing it on the constrained manifold can be bypassed. In
fact, the result of the procedure is known to be a Hamiltonian
system defined on the constrained manifold. The Hamilton's
function of the reduced constrained system is just the original
Hamiltonian evaluated at the constrained manifold \cite{Arnold1}.

\section{Coarse-grained description of the spin}

\subsection{Equivalence of states}

The phase space of the classical system is the sphere denoted
$\Gamma$, of some radius $J$. Pure states of the classical
system are the points of the sphere, parameterized by the spherical
angles $(\theta,\phi)$, or equivalently by the corresponding
vector ${\mathbf J}(\theta,\phi)=(J_x(\theta,\phi),J_y(\theta,\phi),J_z(\theta,\phi))$
of fixed length $|\mathbf{J}(\theta,\phi)|=J$, or by the points of the complex plane
$z=-\tan(\theta/2)\exp(-i\phi)$. The symplectic structure of the classical
phase space is expressed, for example, in terms of $z$ by
\begin{equation}
\omega_{\Gamma}=2i J{dz\wedge d\bar z\over (1+|z|^2)^2}.
\end{equation}
The real coordinates $(q,p)$ given by
\begin{equation}
q+ip={\sqrt {4J}z\over {\sqrt{1+|z|^2}}}
\end{equation}
are the canonical coordinates with respect to $\omega_{\Gamma}$.
The basic variables $J_x(\theta,\phi)$, $J_y(\theta,\phi)$,
$J_z(\theta,\phi)$ form the $su(2)$ algebra $\{J_x,J_y\}_{\Gamma}= J_z$
(and cyclic permutations) with respect to the Poisson bracket induced by (15).
Thus, the dynamical algebra of the classical system is the $su(2)$ algebra.
The Hamilton function of the classical system is not necessarily an
element of the $su(2)$ algebra but is assumed to be expressible as a
simple function of $J_x(\theta,\phi)$, $J_y(\theta,\phi)$, $J_z(\theta,\phi)$.

The quantum system with the same dynamical algebra is the quantum spin.
Cartesian coordinates $\hat J_x$, $\hat J_y$, $\hat J_z$ of the spin
operator $\hat{\mathbf J}$ satisfy the $su(2)$ commutation relations:
$[\hat J_x,\hat J_y]=i\hat J_z$ (and cyclic permutations).
The Hilbert space of the spin of size $J$ is the space of
$(2J+1)$-dimensional irreducible $su(2)$ representation.

$su(2)$ coherent states $|\Omega\rangle$ can be defined as the quantum states that
minimize the total quantum fluctuation of $\hat{\mathbf J}$ \cite{Delburgo}:
$\Delta^2_{\psi}\hat J_x+\Delta^2_{\psi}\hat J_y+\Delta^2_{\psi}\hat J_z$.
Thus, $\Omega$ is the coherent state iff
\begin{equation}
\Delta^2_{\Omega}\hat J_x+\Delta^2_{\Omega}\hat J_y+\Delta^2_{\Omega}\hat J_z=J.
\end{equation}
This is the main property of the coherent states for our purposes.
Alternatively, the coherent states are defined as the eigenstates
corresponding to the maximal eigenvalue of the operator
\begin{equation}
(\hat J_x \sin{\theta}\cos{\phi}+\hat J_y \sin{\theta}\sin{\phi}
+\hat J_z\cos{\theta})|\Omega(\theta,\phi)\rangle=J|\Omega(\theta,\phi)\rangle.
\end{equation}
The set of coherent states is parameterized by the points of the
two-dimensional spherical submanifold $\Gamma$ of the
$2(2J+1)$-dimensional quantum phase space ${\cal M}$. Furthermore
each coherent state satisfies:
$J_x^2(\Omega)+J_y^2(\Omega)+J_z^2(\Omega)=J^2$, where
$J_i(\Omega)\equiv\langle\Omega|\hat J_i|\Omega\rangle$.
Thus the coherent states are in a one-to-one relation with the points
of the phase space of the classical system.

A classical pure state $(\theta,\phi)$ does not make a distinction
between pure quantum states such that the average of the vector operator
$\hat{\mathbf J}$ is a vector collinear with $\mathbf{J}(\theta,\phi)$.
Thus, we define an equivalence relation on ${\cal M}$ (i.e. on
${\cal H}$) as follows:
\begin{equation}
X_1\sim X_2 \quad {\rm iff}\quad
\mathbf{J}(X_1)=\kappa\:\!\mathbf{J}(X_2),
\end{equation}
where $\kappa$ is a positive scalar. The two quantum states are
equivalent if they give collinear expectation vectors
$\langle\hat{\mathbf J}\rangle_X=\mathbf{J}(X)\equiv (J_x(X),J_y(X),J_z(X))$.

Each equivalence class of quantum pure states $[X]$ contains one
and only one coherent state, i.e.\ an $\Omega_X\sim X$ such that
$J_x^2(\Omega_X)+J_y^2(\Omega_X)+J_z^2(\Omega_X)=J^2$. The
partition of ${\cal M}$ by the equivalence relation $\sim$
represents the coarse-grained description of the space of quantum
pure states. The coarse-grained quantum states, i.e.\ the coherent
states, are parameterized by the classical pure states.

\subsection{Classical dynamics: Constraining the quantum dynamics}

Schr\"odinger evolution equation for $\psi(t)$, or its Hamiltonian
form for $X_{\psi}(t)$, does not preserve the equivalence classes of
quantum states (19) and the manifold of coherent states is not invariant.
On the other hand, the system with the same Hamiltonian and additional
constraints introduced in such a way that the manifold of coherent states
is invariant also preserves the equivalence classes of quantum states.
This constrained Hamiltonian system when restricted on the manifold of
coherent states generates by definition the dynamics of the coarse-grained
reduced quantum system. The constrained evolution of the quantum system
with the corresponding total Hamiltonian preserves small the total quantum
fluctuation of $\hat{\mathbf J}$ for all times, which is the crucial
property in the analyzes of its macro-limit.  It is our goal in this
subsection to construct the constrained Hamiltonian evolution
such that the manifold of coherent states is preserved.
Due to the unique representation of the equivalence
classes by coherent states this condition on the evolution also
implies that the coarse-grained states, i.e.\ the equivalence classes,
evolve as a single unit.

The manifold of coherent states $\Gamma$ is uniquely determined by the
total quantum fluctuation of the basic operators $\hat J_x,\hat J_y,\hat J_z$
which is minimal if and only if the state is a coherent state.
Thus, the constraint
\begin{equation}
\Phi(X)=\Delta^2_{X}\hat J_x+\Delta^2_{X}\hat J_y+\Delta^2_{X}\hat J_z-J=0
\end{equation}
determines the sphere of coherent states. The evolution of the coarse-grained
system is defined to be the constrained Hamiltonian evolution with the
given Hamiltonian $H(X)$ and the constraint (20).

The master constraint that we want to fulfill is given by the function (20),
but that constraint is not regular because it is equivalent with
$J_x^2(X)+J_y^2(X)+J_z^2(X)-J^2=0$. Similarly to the case of oscillators
treated in \cite{us}, the application of Dirac's procedure with this
constraint as the initial primary one is not straightforward, and would
imply an additional number of secondary constraints.
However, since the constrained manifold $\Gamma$ is known to be symplectic,
the constrained system reduced on $\Gamma$ is Hamiltonian with the Hamilton's
function given simply by $H(\Omega)=\langle\Omega|\hat H|\Omega\rangle$.

Alternatively, one could obtain the constrained evolution equations on the full
space ${\cal M}$, using an appropriate form of the primary constraints and
then reduce the constrained system on the constrained manifold. Following the
idea presented in \cite{us}, one should replaced the irregular primary constraint
(20) by a more effective equivalent primary constraint, formulated using the
equivalence of states. The primary constraint that should be imposed would require
that the average of the Hamiltonian $H(X)$ is equal to its average in the equivalent
coherent state $\Omega_X$
\begin{equation}
\Phi(X)=H(X)-H(\Omega_X)=0.
\end{equation}
The Lagrange multiplier in the total Hamiltonian $H_{tot}(X)=H(X)-\lambda\Phi(X)$
is simply $\lambda=1$ and thus the total Hamiltonian is
\begin{equation}
 H_{tot}(X)=H(\Omega_X).
\end{equation}
As already stated, the restriction of the evolution of the resulting constrained
system onto the constrained manifold of coherent states is guided
simply by the Hamiltonian $H(\Omega)$.

In summary, the Hamilton's function of the coarse-grained and
reduced system is just the $\langle\Omega|\hat H|\Omega\rangle$.
The states of the coarse-grained system are equivalence classes
represented by the coherent states. The quantum constrained dynamics
preserves the equivalence classes and the total quantum fluctuation
remain minimal (but nonzero) throughout the evolution. Of course,
the coarse-grained system is not classical. The coherent states have
nonzero overlap and the quantum fluctuations are, although minimal,
nonzero. For example, if the quantum Hamiltonian is a nonlinear
expression in terms of the basic operators $\hat J_x,\hat J_y,\hat J_z$
its expectation $\langle\Omega|\hat H(\hat J_x,\hat J_y,\hat J_z)|\Omega\rangle$
is different from $H(\langle\Omega|\hat J_x|\Omega\rangle,\langle\Omega|\hat
J_y|\Omega\rangle,\langle\Omega|\hat J_z|\Omega\rangle)$ where $H(\dots\/)$
is of the same form as $\hat H(\dots\/)$. Due to the constraint dynamics of
the coarse-grained model, the total quantum fluctuations are minimal and
the difference between those two expressions is all the time bounded by terms
of the leading order $1/J$. The difference becomes arbitrary small as $J$
becomes sufficiently large. Macro-limit of the coarse-grained system is
discussed in the next section.

Notice that the reduced constrained Hamilton's function
$H(\Omega)= \langle\hat H\rangle_\Omega$ is a valid Hamiltonian
for any Hermitian operator $\hat H$, while the Hamilton's function
of the classical model must be a function of the expectations of
the basic operators $\langle\hat J_x\rangle_\Omega,\langle\hat
J_y\rangle_\Omega, \langle\hat J_z\rangle_\Omega$.

Sometimes the Hamiltonian of a quantum system is a sum of terms
linear in the dynamical algebra generators and a small
perturbation containing some nonlinear terms. In such cases the
manifold of coherent states of the dynamical algebra is
approximately invariant over some finite time. This fact has been
used (see for example \cite{kinez} or more recent
\cite{Korsh1,Korch2}) to propose and study an approximation of the
exact quantum dynamics, from some coherent state or a mixture of
such, by the equations of classical Hamiltonian form with the
Hamilton's function given by $\langle\hat H\rangle_\Omega$. In our
approach the reduced constrained Hamiltonian equations of the
coarse-grained system appear as the evolution equations of the
equivalence classes of the quantum states. The equations are the
evolution equations of the coarse-grained quantum system whatever
the form of the Hamilton's operator is.

\section{Macro-limit of the reduced constrained system and the classical model}

In this section we want to demonstrate that for sufficiently large
$J$ the states of the coarse-grained system and their evolution
display properties of a classical system. In particular we shall
show that for sufficiently large $J$: (a) the coarse-grained
system appears to be in one and only one classical state; (b) the
state of the coarse-grained system can be determined without
measurable disturbance; (c) the evolution of the coarse-grained
system is such that the property (a) is valid for all times, or in
other words, ratios $\Delta\hat J_i/ \langle\hat J_i\rangle$
$(i=x,y,z)$ remain arbitrary small during the evolution; (d) the
evolutions of the coarse-grained system and of the classical model
become indistinguishable. As the consequence of these properties
the reduced constrained system with sufficiently large $J$ is the
same as the classical model for all observational devices with
arbitrary but finite accuracy.

The properties (a), (b) and (c) are based on the fact that the overlap
between the coherent states $|\langle\Omega|\Omega'\rangle|=\cos^{2J}(\alpha/2)$
becomes arbitrary small for sufficiently large $J$, where $\alpha$ is the angle
between directions $(\theta,\phi)$ and $(\theta',\phi')$ correspondong to
$\Omega$ and $\Omega'$, respectively. Thus, for an observation with arbitrary
but finite accuracy, different coherent states appear as orthogonal.

The property (a) is in fact the same as the near orthogonality of
the coherent states for large $J$. To demonstrate the property (b)
consider the measurement in the overcomplete basis  given by the coherent states.
Upon such measurement the representative of the coarse-grained
state $|\Omega\rangle$ is transformed into $|\Omega'\rangle=\int
d\Omega''|\Omega''\rangle\langle\Omega''|\Omega\rangle$ which is
approximately equal to $|\Omega\rangle$ due to near orthogonality
of the coherent states for large $J$. Thus, disturbance of the states
of the reduced system with sufficiently large $J$ by the measurement
of classical properties is negligibly small.

Evolution of the reduced system is defined precisely such that the
dispersions $\Delta\hat J_i$ $(i=x,y,z)$ remain small, and thus
the ratios $\Delta\hat J_i/\langle\hat J_i\rangle$ are arbitrarily
small for sufficiently large $J$. This implies in particular that
$\langle\Omega|f(\hat J_x,\hat J_y,\hat
J_z)|\Omega\rangle=f(\langle\Omega|\hat
J_x|\Omega\rangle,\langle\Omega|\hat
J_y|\Omega\rangle,\langle\Omega|\hat J_z|\Omega\rangle)+{\mathcal
O}(1/J)$ where $f$ is an arbitrary polynomial expression. Thus,
the evolution of the coarse-grained and reduced system with large
$J$ and the Hamiltonian $\langle\Omega|\hat H(\hat J_x,\hat
J_y,\hat J_z)|\Omega\rangle$ is indistinguishable from the
classical evolution generated by the Hamilton's function
$H(\langle\Omega|\hat J_x|\Omega\rangle,\langle\Omega|\hat
J_y|\Omega\rangle, \langle\Omega|\hat J_z|\Omega\rangle)$. We can
conclude that the reduced constrained system for large $J$
displays all typical properties of a classical Hamiltonian system
in all physically possible observations.

\section{An example}

If the Hamiltonian of the quantum system is a nonlinear expression
of the basic operators of the dynamical algebra then the
Schr\"odinger evolution with the Hamiltonian $\hat H(\hat J_x,\hat
J_y,\hat J_z)$, the evolution of the reduced constrained system by
the Hamiltonian $H_{tot}$ (22), and the evolution of the classical
model with the Hamiltonian $H(\langle\hat J_x\rangle,\langle\hat
J_y\rangle,\langle\hat J_z\rangle)$, starting from the same
coherent state, are all different. However, when the spin size $J$
increases the difference between the evolution of the
coarse-grained and reduced system and of the classical model
decreases. Thus, for sufficiently large $J$ the differences become
negligible over arbitrary large periods of time. These facts are
illustrated in this section using as an example the following
Hamiltonian:
\begin{equation}
\hat H=\epsilon \hat J_z -\lambda \hat J_x + \mu\hat J_z^2
\end{equation}
where $\epsilon, \lambda$ and $\mu$ are parameters. The Hamiltonian
(23) appears as the most convenient form of the two mode
Bose-Hubbard model \cite{Korch2}.

The constrained system reduced on $\Gamma$ is a Hamiltonian system
on $\Gamma$ with the Hamilton's function given by $\langle\hat H\rangle_\Omega$.
To express it as a function of the canonical coordinates $(p,q)$ of
$\Gamma$ we need the appropriate expectations of linear and quadratic
operators in terms of $(p,q)$. The relevant formulas are given by:
\begin{eqnarray}
\langle\hat J_x\rangle(p,q)&=&{q\over 2}(4J-q^2-p^2)^{1/2},\nonumber\\
\langle\hat J_y\rangle(p,q)&=&-{p\over 2}(4J-q^2-p^2)^{1/2},\nonumber\\
\langle\hat J_z\rangle(p,q)&=&{1\over 2}(q^2+p^2-2J),
\end{eqnarray}
and
\begin{equation}
\langle\hat J_z^2\rangle=\langle\hat J_z\rangle^2(p,q)+
\frac{1}{8J}(p^2+q^2)(4J-p^2-q^2).
\end {equation}
The last term in (25), proportional to $1/J$ represent the quantum
correction to the expectation of the nonlinear operator $\hat J_z^2$.

The Hamilton's function $H(p,q)\equiv H(\Omega(p,q))$ of the constrained system
reduced on $\Gamma$ is given by
\begin{eqnarray}
H(p,q)&=&{\epsilon\over 2}(q^2+p^2-2J)-\lambda{q\over 2}(2J-p^2-q^2)^{1/2}\nonumber\\
&+&\mu\left[{1\over 4}(q^2+p^2-2J)^2+{1\over 8J}(p^2+q^2)(4J-p^2-q^2)\right].
\end{eqnarray}

The Hamilton's function of the classical model is obtained by
assuming that the last term in square brackets in (26) is in fact equal to zero,
i.e.\ $\langle\Omega|\hat J_z^2|\Omega\rangle=\langle\Omega|\hat J_z^2|\Omega\rangle^2$.
The classical Hamiltonian reads
\begin{eqnarray}
H_{cl}(p,q)&=&{\epsilon\over 2}(q^2+p^2-2J)-\lambda{q\over 2}(2J-p^2-q^2)^{1/2}\nonumber\\
&+&{\mu\over 4}(q^2+p^2-2J)^2.
\end{eqnarray}

\begin{figure}[t]
\includegraphics[width=0.8\textwidth]{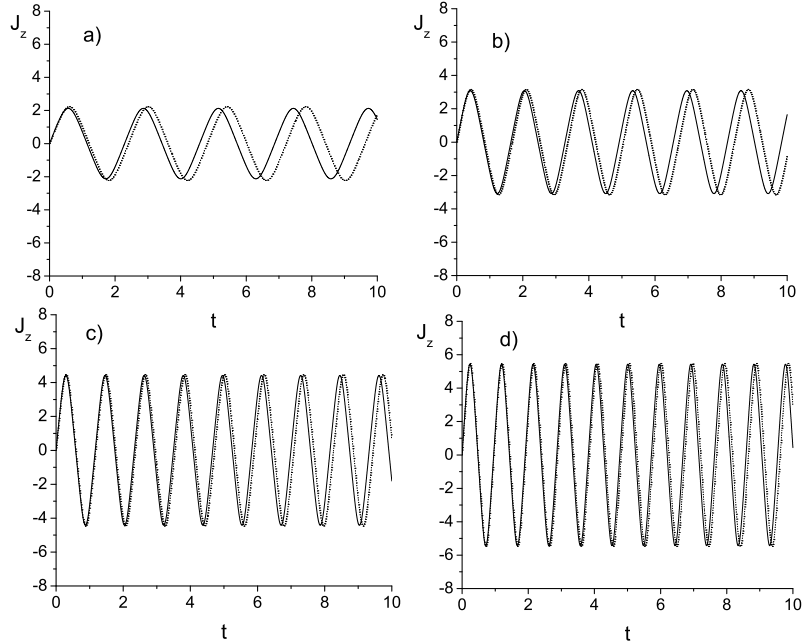}
\caption{Illustrates constrained (dotted line) and classical
(thick line) evolution for a) $J=5$, b) $J=10$, c) $J=20$ and d)
$J=30$. Time series $J_z(t)=\langle\Omega(t)|\hat J_z|\Omega(t)\rangle$
are shown. All quantities are dimensionless.  Parameters are $\lambda=\mu=1,\epsilon=0$ and the initial
state is the coherent state $|\Omega_0\rangle=(\theta=\pi/2,\phi=0)$.
\label{Fig1}}
\end{figure}
In Fig.\ 1 we illustrate the evolution of the average $\langle\hat J_z\rangle$
generated by the  constrained system (26) and by the classical model (27).
The initial state for the constrained system is the coherent state
$(\theta=\pi/2,\phi=0)$ implying $(p_0=\sqrt{2J},q_0=0)$.
The main conclusion of Fig.\ 1 is that the constrained and the classical
evolutions become indistinguishable as $J$ becomes sufficiently large.

\section{Discussion and Summary}

We would now like to compare the coarse-graining whose
fundamental role is analyzed in this paper with two different
types of coarse-graining commonly used in the studies of
micro-macro transition. We consider the coarse-graining (a) in the
mean field type approach to the appearance of macro-properties
and (b) motivated by the finite precision of the measuring
instruments. Let us first discuss the coarse-graining of the type
(a). Typical example is the treatment of macroscopic magnetization
defined as the average of the spin components over an ensemble of
spins: $\hat J_i=\sum_k\hat\sigma_i^k$ $(i=x,y,z)$. All states of
the ensemble which give the same average of the large spin $\hat
{\mathbf J}$ components are considered equivalent. However, the
large spin is a quantum system which can be in states corresponding
to superpositions of macroscopically distinct eigenvalues of its
observables. No classical behavior is implied by the
coarse-graining that replaced the ensemble of spins by the single
large spin \cite{Brukner2,jaPRA}. Furthermore, the
coarse-graining that treats as indistingushable the eigenstates of
an observable with nearby eigenvalues also does not introduce the
classical behavior. Thus the coarse-graining of type (a) or of type
(b) although justified are not crucial for the explanation of the
emergence of classical systems.

To summarize, we have analyzed the conditions for the emergence of
a classical Hamiltonian dynamical system on the sphere from the quantum
spin. The main condition behind the classical appearance is that only a
limited set of averages of quantum observables is distinguished in
the classical system. This naturally leads to the corresponding
equivalence relation among the quantum states.
The equivalence relation represent a type of coarse-graining,
such that in each equivalence class there is one and only one
state with minimal total quantum fluctuation, i.e.\ the
corresponding coherent state. If the equivalence classes or
their representatives are to be identified with states of the
appropriate classical system then they must evolve as single
units. This leads to the constrained Hamiltonian dynamics which
preserves the manifold of coherent states. For an observer with
sufficiently limited observational accuracy the constrained system
reduced on the manifold of coherent states displays all
characteristic properties of a classical system. The upper limit
on the observational accuracy with which the system appears as
classical can be increased as the size of the spin $J$ is
increased but is ultimately limited by the physical nature of
possible observational devices.

In the reference \cite{us} we have applied the same ideas and
methods to explain the emergence of the classical system of
oscillators from the corresponding quantum system. The
important difference, of ultimately geometrical origin, between
the system of oscillators and the spin is in the form of the
equivalence relation i.e.\ the form of the coarse-graining.
Nevertheless, in both cases the role of the coarse-graining is the same.
The two examples of the oscillators and the spin suggest a general
explanation of the emergence of classical models from quantum systems
with the same dynamical Lie algebra.

\begin{acknowledgments}
This work is partly supported by the Ministry of Education and
Science of the Republic of Serbia, under Grants No.\ 171017,
No.\ 171028, No.\ 171038, and No.\ III45016. We would like to
acknowledge useful comments from Prof.\ \v{C}.\ Brukner,
Prof.\ M.\ Everitt and dr E-M. Graefe.
\end{acknowledgments}

\end{document}